\documentclass[conference]{IEEEtran}
\usepackage{latexsym}
\usepackage{graphicx}
\usepackage{mathptmx}
\usepackage{amsmath}
\usepackage{amsfonts}
\usepackage{amssymb}
\usepackage{amsbsy}
\usepackage{amsthm}
\usepackage{array}
\usepackage{multirow}
\usepackage{hhline}		
\usepackage{subfigure}
\usepackage{mathptmx}
\usepackage{mathtools}
\usepackage{todonotes}

\usepackage{enumitem} 
\usepackage[square,sort&compress,comma,numbers]{natbib}

\usepackage{booktabs} 
\usepackage{float}
\usepackage{scrhack}
\usepackage{listings}
\usepackage{color}
\usepackage{pgfplots}
\usepackage{tikz}
\usepackage{cleveref}
\usepackage{here}
\usepackage{url}

\pgfplotsset{ 
	compat=1.14, 
	/pgf/number format/.cd, 
	fixed, 
	set thousands separator = $.$,
	min exponent for 1000 sep=4 
}

\ifCLASSINFOpdf
\else
\fi
\begin{document}

\title{\vspace*{-0.1cm}Optimized Travel to Meetings on a Common Location of Geographical Distributed Participants}
\author{
	\IEEEauthorblockN{Peter Hillmann, Bastian K\"uhnel, Tobias Uhlig, Gabi Dreo Rodosek, and Oliver Rose}
	\IEEEauthorblockA{Universit\"at der Bundeswehr M\"unchen\\
		Neubiberg, 85577, GERMANY\\
		Email: \{peter.hillmann, bastian.kuehnel, tobias.uhlig, gabi.dreo, oliver.rose\}@unibw.de \vspace*{-0.2cm}
	}
}


\maketitle

\begin{abstract}
Members of international organizations often meet in person at a common location for discussions.
There is frequently disagreement over the place and time of the meeting due to the different travel efforts of the members.
They usually travel by plane and their travel expenses depend on the flight connections.
This paper presents an approach to calculate the optimized location and time, where and when distributed partners should meet.
The presented system considers the requirements and specifications of each individual member.
It respects earliest starting time of an event and non night flights.
The optimized result is evaluated with regard to multiple objectives.
We focus on the minimization of costs and travel time.
Our search algorithm identifies individual travel data for all members for a potential event.
The output provides recommendations for the global best appointments and offers further information for the partners.
Our system saves expenses and time for all members and allows adjustment as well as compensation.
\end{abstract}


\IEEEpeerreviewmaketitle

\section{Motivation}
Cost pressures and time efficiency are main reasons to improve or change existing procedures and solutions.
In 2014, the European Union spent 180 million Euro on the travel expenses of its deputies in order to hold the monthly parliamentary session in Strasbourg instead of Brussels \cite{Mendick2014}.
Furthermore it is hardly reasonable to accept travel times of more than the double meeting time, especially if the meeting takes only a few hours.
Reasons for such circumstances are venues where no time-saving travel connections exist.
The main aspect of the business traveller is the relationship between travel time, useful time, and costs.
So, they try to avoid additional overnight stays.
Nevertheless, it is sometimes unavoidable because of unavailable travel connections or temporally impractical trips.

Consequently, an adequate planning is necessary to save expenses and time.
The aspiration is a practical application to obtain a usable system for daily use.
We present an approach to organize meetings on an optimized place and at a perfect time at a common location. 
Our system considers the conveniences of the members and legal regulations.
We focus on results corresponding to global optimum to avoid discussions between the participants of the meeting.
It uses public available information of airlines for flight connections to calculate favourable venues.
Our system is evaluated by solving the typical scenario of the European Union Agency for Network and Information Security (ENISA).
We demonstrate that a usage of our system leads to cost and time improvements for such organizations and situations.
Another use case for our system is the application area of logistics.
It also can be used for the identification of consolidation depots including time schedules for timely transportation of goods.
This supports a just in time delivery with reduced transportation costs and leads to a higher quality of service for the customers \cite{7136635}.

The remainder of this paper is structured as follows. 
In Section 2, we describe a typical scenario and requirements for the problem.
Section 3 briefly discusses related work in the application area.
The main part in Section 4 presents the calculation system and particular specifications.
Thereafter, Section 5 shows some experiments and practical usage, before Section 6 concludes this paper.\\

\IEEEpubidadjcol 
\section{Scenario}\label{sec:scenario}
\begin{figure}[H]{
		\centering
		\includegraphics[width=0.48\textwidth]{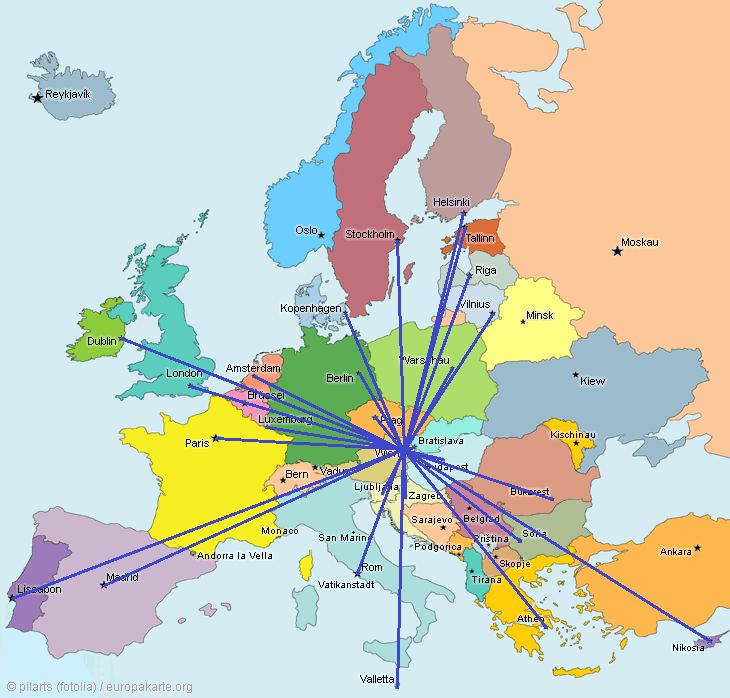}
		\vspace{-0.1cm}
		\caption{Scenario of the Problem with venue in Vienna \cite{Molkenthin2016}}
		\label{fig:Szenario}
}\end{figure}

Members of an organization in Europe has to meet regularly for personal discussions at a common location.
The events are usually planned at a participants place, because of the local organization.
Nevertheless, this is not mandatory. 
The problem is that the partners of the European Union (EU) institutions are spread across all 28 member states in Europe.
The scenario is outlined in \mbox{Figure \ref{fig:Szenario}.}
\vspace{-0.3cm}

The travel distances between the cities are long. 
Therefore, the representatives travel mainly by plane due to shorter travel time as the costs are covered by their countries.
Usually, a business traveller intends to have a round trip \cite{KagelmannHahn1993}.
The considered time window of the meeting is typically specified within one or two weeks.
The time point can be determined flexibly within the work days.
The length of the meetings usually varies between half a day up to three days.
Nevertheless, the design of the planning system has to be flexible in all the aforementioned mentioned aspects. 
Currently, a representative proposes to organize the next meeting at his place, because the member has a room and catering available.
This start point of planning does not lead to either a time or cost optimization for the travelling participants and their organisations.
Furthermore, this probably results in long flight times, multiple transfers and unnecessary overnight stays.
However, a venue and time for the event has to be previously defined.

For a pleasant meeting and organization, the following requirements have to be considered by our planning system:\\

\begin{itemize}
\item Flexible specification of the planned meeting duration 
\item Maximum length of a session term on one day
\item Earliest time of start and latest time of end of a meeting
\item A meeting never starts e.g. on Saturday or Sunday
\item Earliest travel time to the meeting on the starting day
\item Latest time of arrival on a day before (overnight stay)
\item Latest time of departure or latest time of arrival at destination after a meeting (overnight stay)\\
\end{itemize}

For simplification, we assume that all participants travel by plane.
This reduces the initial amount of data sources for calculation and demonstration.
Other data sources of further transportation systems and vendors like train or car can be connected to the system in the same way.
Nevertheless, the approach is not specific for flight connections and works also for a wider range of travel information.
Therefore, the structure of the processed data sources has to be designed generic.
The result has to include information about the location of the meeting as well as starting time and end time.
Beside this, the system has to provide the flight data to and from the venue individually for every participant.
For an enhanced version of the problem, the travel time and costs of the individual participants should be compensated through multiple meetings.\\

\section{Related work}
To the best of our knowledge and research, we realized that only a few published solutions for this typical problem exists. 
For short term meetings exists a solution \cite{5319235}, in which all participants are not widely spread.
Other current approaches with distributed members focus only on one part of the problem and thereby only on one criteria.
Their focus is either an optimized location selection based on distances \cite{Kleinberg2013} or time fitting of multiple overlapping calendars \cite{7791872}.

The main application area is the navigation. It purpose is to lead the user on the fastest track to the destination \cite{Siemoneit2015}.
However, the destination is unclear at the beginning.

A simple solution of the location problem is the calculation of the center point from all locations \cite{GeoMidpoint2007}.
Thereby the participants meet in the center whereby they have a reduced distance to travel.
This is comparable with the k-center or k-median problem with one center which can be solved by clustering approaches \cite{Chaudhuri1996}.
Nevertheless, it does not consider flight schedules.

Alternatively, the location and time can be defined alternating.
This leads to an equal balance between the participants, which is obviously not an optimized solution.
Based on the assumption that the largest airport offers the best flight connections, the venue can be defined at this city.
This supports a reduced time overhead for all participants before or after a meeting.
But this approach also does not consider the problem in detail.

Alternatively, many solutions suggest a web based conference by voice and video call.
But this is not the same like face-to-face meetings and for special meetings with security clearance not possible.

\section{Concept}
\begin{figure}[hbtp]{
		\centering
		\includegraphics[width=0.48\textwidth]{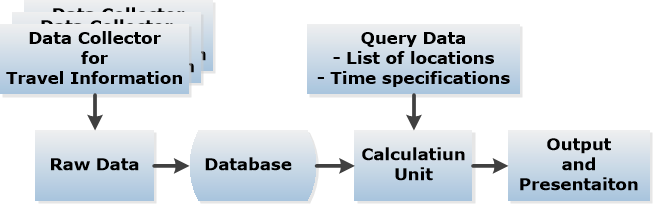}
		\caption{Overview about the calculation system}
		\label{fig:Prozess}
}\end{figure}

Our proposed system is separated in multiple components, which are dedicated for a specific task.
The Figure \ref{fig:Prozess} provides an overview of the system components and the main process steps.
The system starts with a data collector unit to search and query travel information.
The raw data are converted to a common and generic data format.
Subsequently the information is stored in a database fitting to a general data format of travel and connection information.
These data are used by the calculation unit to identify a solution for the meeting problem.
It uses the specified input information from a query form requested by the user.
The solution is stored in a separate table of the database with all flight connections of the participants.
In the following, each step is explained in detail.
Thereby the challenges are mentioned and adequate solutions are presented.

\subsection{Collecting data}
For every source, the data collector needs to be individually realized, because of the specific interfaces of the data provider.
So the system has multiple data collectors filling the database with information. 
Airlines or public transport provide their schedules via website, which can be usually queried through an API.
For this, often a token needs to be generated to send authorized requests to the external data provider.
A controller handles all these authentication credentials and manages the data collectors.
Further sources of travel information are given by PDF or search engines and can be parsed or queried separately.
These requested data have different types of formats.
For that reason, the raw data are converted to a standardized format of XML.
This file format has a specific defined structure according to the generic database and mandatory information.
Missing information are filled up with default values whereas wrong data are filtered out.
The data of the several data collectors are stored into the database.
Based on our scenario, we use flight schedules mainly from \mbox{Lufthansa (LH).}

\subsection{Database}
All travel information from the common XML structure are stored in a MySQL database.
The data need to be queried beforehand to speed up the calculation process.
Furthermore, this step logically decouples the necessary process steps and allows an offline usage of the system.
Some portals for travel information have time-based query limits, which can be bypassed with an own database.
However, the stored information in the database need to be updated regularly.
Therefore an updating component exists.
To handle many requests, the system uses an In-Memory database to avoid bottlenecks.

For every location and city, the database has a separate table with the following entries:\\

\begin{itemize}
\item Departure location (Unique abbreviation)
\item Time of departure (Local time)
\item Destination location
\item Time of arrival (Local time)
\item Duration of transfer\\ (To overcome problems with different time zones)
\item External reference code of the transfer
\item External ID for booking
\item Number of stopover
\item Day of travel\\ (Encoded with 1 to 7 for Monday to Sunday)
\item Cost\\
\end{itemize}

So far, we decided to forgo on the normalization of the table structure to make the requests simpler and transparent.
To fill the database with mandatory information, the list of locations form all meeting participants has to be known beforehand.
Depending on the list, the data collector queries the information.

For the solution, two additional tables are created to store the information of the calculation unit in the database.
The first table \textit{res\_member} contains all data for every individual participant and has the following entries:\\

\begin{itemize}
\item Location of the participant
\item Possible venue (Manual or automatic detection \cite{Francis1991})
\item Sum of travel time for departure and arrival
\item Number of stop over
\item Sum of costs\\
\end{itemize}

The second table \textit{res\_meeting} contains the overall result of all participants at a specific venue.
It contains the following information:\\

\begin{itemize}
\item Location of the meeting
\item Time of meeting start
\item Time of meeting end
\item Sum of travel time of all participants
\item Travel time of the participant with the longest journey
\item Sum of costs of all participants\\
\end{itemize}

\subsection{Calculation Unit}
Our focus for a solution is the global optimum to avoid discussions between the members.
So the calculation unit has to use a complete search algorithm to determine the location and time that is most suitable.
Therefore, no heuristic approaches or Linear Programming can be used to achieve this mandatory solution quality.
The system takes the input information from the query form given to the user of the planning system.
This includes information about the required length of the meeting in hours as well as further specification listed in Section \ref{sec:scenario} for a pleasant trip.
The algorithm identifies a location and time which corresponds to the search criterion, using the flight data stored in the database.
It consider specific requirements like the earliest start of the meeting.

Our strategic search is based on the time of the meeting.
It is the decisive element of this complex problem.
So the meeting is the starting point of the search.
All further steps are based on the planned starting time of the meeting.
We vary the meeting time from earliest possible to latest possible to fulfill the search.
As the time is a continuous search space, we discretized it in time slices.

\begin{figure}[hbtp]{
		\centering
		\includegraphics[width=0.48\textwidth]{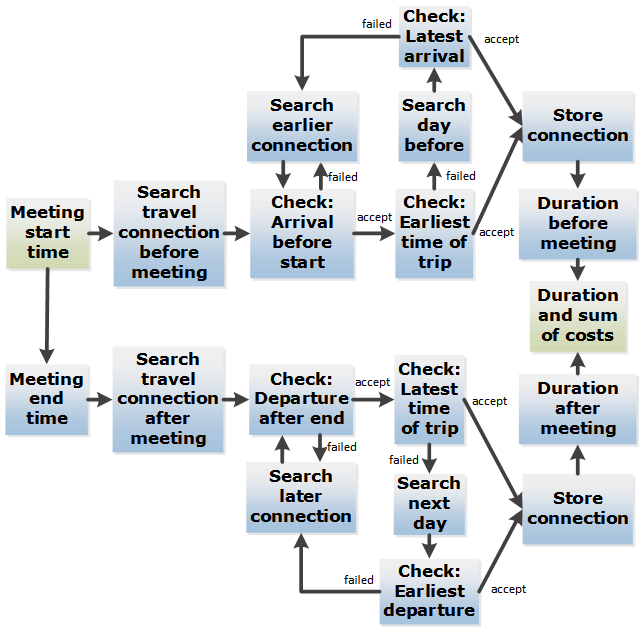}
		\caption{Finding travel connection considering specified requirements}
		\label{fig:Algorithmus}
}\end{figure}

The complete search algorithm uses multiple nested loops.
First of all, the outer loop iterates over all possible days for a meeting.
The next inner loop iterates over the time on a specific day.
The time is sampled in ten minutes time slots to reduce the search space.
Nevertheless, it can flexible specified.
The considered time point defines the start of the planned meeting.
This information triggers the last nested loop about the location.
The algorithm iterates through all possible venues for a specific time and calculates for every participant the travel time and costs.
The calculation of the individual travel data is separated in five parts.
These are the outward trip, meeting time with additional overhead before and after the meeting, and the return trip, see Equation \ref{eq:timesum}.

\begin{equation}
t_{sum} = t_{out} + t_{before} + t_{meeting} + t_{after}+ t_{return}
 \label{eq:timesum}
\end{equation}
\newline
The overall process of travel calculation is shown in Figure \ref{fig:Algorithmus}.
It presents the individual and particular steps for the outward trip and the return trip.
For each part of the journey, the algorithm searches in the database the best suitable travel connections considering the specified constraints.
For the outward trip, the algorithm moves backward from the starting time and search for the first connection, which fulfills all conditions.
Accordingly, the algorithm moves forward form the ending time of the meeting to identify a connection for the return trip.
The results are stored in the table \textit{res\_member} of the database.
Subsequently, the algorithm summarize the solution and stores the result in the table \textit{res\_meeting} of the database.

The ending time of the meeting is calculated depending on the checked starting time and the specified duration.
Afterwards, two parallel processes are started, once for the outwards trip and once for the return trip.
For each one-way trip a fitting travel connection is searched in the database.

As the data are stored in a database, the SQL query for the travel information is defined in a specific way so that the necessary result is returned.
This avoids algorithmic search in the return data of the database and speed up the process.
Even if it is a full search, the amount of calculation is small enough to finish it in a few minutes depending on the amount of different locations.
Most time consuming part is the request of external data sources.
In this way, we obtain the global optimum instead of another solution comparable to the current situation presented in Section \ref{sec:scenario}.

\subsection{Output}
The last component of the system evaluate the calculated solutions with respect to the specified criteria.
There are two main criteria, which are of practical relevance:\\

\begin{itemize}
\item Minimized sum of travel time or costs
\item Minimized maximum travel time with reduced stop over\\
\end{itemize}

Therefore the results in table \textit{res\_meeting} are compared with each other.
Our algorithm iterates over the table.
For same locations, the earlier one on this venue is used.
The best results are shown with the overall properties according to the criteria. 
Multiple solutions at different locations are presented to give the user of the system different choices.
Every meeting solution include the information for every participant, i. e. which travel connection should be used.
Furthermore, the location as well as the starting and ending time of the meeting is provided.\\

\section{Experiments}
We made multiple experiments to demonstrate the functionality of our system.
Especially the following three evaluations show the improvement by using the system.

\textbf{Experiment 1: }
First of all, the list of locations is reduced to five cities for manual reproducibility: Amsterdam (AMS), Athen (ATH), Brussels (BRU), Frankfurt (FRA), and Stockholm (ARN).
In this example, the meeting is schedules for 6 hours in Frankfurt between 9:20 am and 4:20 pm.
The Table \ref{tab:sumexp1} and \ref{tab:sumexp2} show the individual travel information for the different participants location.

\begin{table} [htbp]
\begin{center} 
\caption{Outwards trips for the different member locations}
\label{tab:sumexp1}
\begin{tabular}{|c|c|c|c|c|}
\hline
Place & Day & Time departure & Time arrival & Type\\ \hline
\hhline{-----}
AMS & Monday & 07:20 am & 08:25 am & Airline LH\\ \hline
ARN & Monday & 06:45 am & 09:00 pm & Airline LH\\ \hline
ATH & Monday & 07:10 am & 09:15 pm & Airline LH\\ \hline
BRU & Monday & 07:05 am & 08:05 pm & Airline LH\\ \hline
\end{tabular}
\end{center}
\vspace*{-0.3cm}
\end{table}

\begin{table} [htbp]
\begin{center} 
\caption{Return trips to the different member locations}
\label{tab:sumexp2}
\begin{tabular}{|c|c|c|c|c|}
\hline
Place & Day & Time departure & Time arrival & Type\\ \hline
\hhline{-----}
AMS & Monday & 04:30 pm & 05:45 pm & Airline LH\\ \hline
ARN & Monday & 05:30 pm & 07:35 pm & Airline LH\\ \hline
ATH & Monday & 06:00 pm & 09:50 pm & Airline LH\\ \hline
BRU & Monday & 04:25 pm & 05:25 pm & Airline LH\\ \hline
\end{tabular}
\end{center}
\end{table}

\textbf{Experiment 2: }
According to our initial scenario in \mbox{Figure \ref{fig:Szenario}}, we calculated for the venue for a 6 hour meeting.
The Table \ref{tab:sumexp} show the bests results.
The column sum represents the overall time effort in minutes.

\begin{table} [htbp]
\begin{center} 
\caption{Best results for a 6 hours meeting of the participants of the EU according to the Scenario in Section \ref{sec:scenario}}
\label{tab:sumexp}
\begin{tabular}{|c|c|c|c|c|c|}
\hline
Venue & Sum & Day start & Time start & Day end & Time end\\ \hline
\hhline{------}
AMS & 80,925 & Monday & 10:00 am & Monday & 4:00 pm\\ \hline
ARN & 88,021 & Monday & 10:00 am & Monday & 4:00 pm\\ \hline
ATH & 91,320 & Monday & 10:00 am & Monday & 4:00 pm\\ \hline
\end{tabular}
\end{center}
\end{table}

\begin{figure}[htbp]
	\centering
	\begin{tikzpicture}
		\begin{axis}[  
   xlabel={Starting time}, 
   ylabel={Duration in minutes}, 
   only marks, 
   ymajorgrids, 
   ymin=30000, 
   ymax=55000,
   ytick distance=5000,
   scaled y ticks = false 
   ]
			\addplot table[x=Beginn, y=Reisezeit] {test.csv};
		\end{axis}
	\end{tikzpicture}
	\caption{Influence of different starting times for a meeting at Frankfurt on the travel duration} 
	\label{fig:variation}
\end{figure}
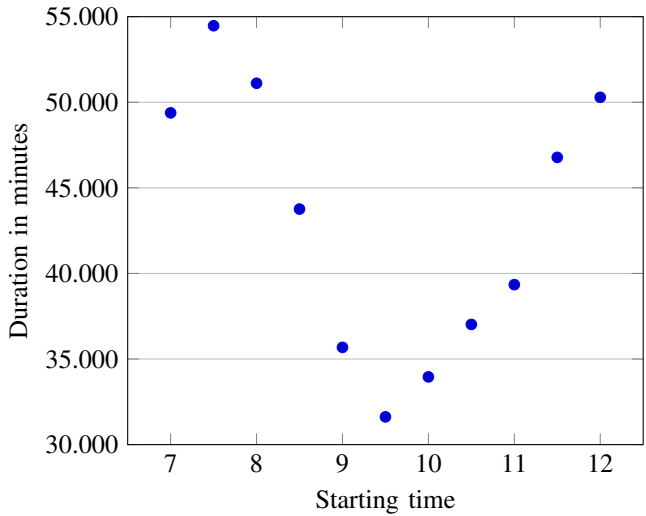

\textbf{Experiment 3: }
To visualize the possible improvements with our approach, we calculate the deviation of time effort for different meeting times on a day.
The Figure \ref{fig:variation} shows the sum of travel time for the location of Frankfurt for a meeting of 6 hours.
It presents the variation between 7 am and 12 pm in 30 minutes steps.
Especially at 9:30, only a few participants need to arrive one day earlier and for others it is not necessary to stay one day longer after the meeting.
This leads to a reduced overall timely effort.
The impact of the starting time of the meeting is so large that the average travel time for the optimum is nearly the half in comparison to the worst case.\\

\section{Conclusions}
The search for an appropriate meeting location and time is a complex challenge.
For practical usage, many side conditions need to be respected.
We presented the concept of a customizable planning system for meetings at a common location of distributed participants.
The design of the approach consists of separate components, which are extensible or flexible replaceable.
It calculates the optimal time to reduce the travel effort and minimizes overhead time considering multiple optimization criteria.
Therefor, the vendor specific data sources are transformed to a generic data format.
As our focus is on the global optimum, we developed a complete search algorithm for the problem using database queries with configurable time slices.
The advantage of this approach is flexible adaptability for extended versions of this meeting problem.
The realized system is a prototype, which will be extend in the future with additional travel information and side conditions.
Furthermore, individual time overheads for local transport will be added.\\


\section*{Acknowledgment}
The authors would like to thank Prof. Dr. Udo Helmbrecht of the ENISA for discussions about the scenario and practical usage.

\bibliographystyle{unsrtnat}
\bibliography{acmart,sigproc} 

\begin{thebibliography}{11}
\providecommand{\natexlab}[1]{#1}
\providecommand{\url}[1]{\texttt{#1}}
\expandafter\ifx\csname urlstyle\endcsname\relax
  \providecommand{\doi}[1]{doi: #1}\else
  \providecommand{\doi}{doi: \begingroup \urlstyle{rm}\Url}\fi

\bibitem[Mendick(2014)]{Mendick2014}
Robert Mendick.
\newblock The farce of the eu travelling circus.
\newblock \emph{The Telegraph}, 2014.

\bibitem[Agrebi et~al.(2015)Agrebi, Abed, and Omri]{7136635}
M.~Agrebi, M.~Abed, and M.~N. Omri.
\newblock Urban distribution centers' location selection's problem: A survey.
\newblock In \emph{International Conference on Advanced Logistics and Transport
  (ICALT)}, pages 246--251, May 2015.

\bibitem[Molkenthin(2016)]{Molkenthin2016}
Marco Molkenthin.
\newblock Die europakarte.
\newblock \emph{Europa}, 2016.
\newblock URL \url{http://www.europakarte.org/}.

\bibitem[Kagelmann and Hahn(1993)]{KagelmannHahn1993}
H.~J. Kagelmann and H.~Hahn.
\newblock \emph{Tourismuspsychologie und Tourismussoziologie}.
\newblock 1993.

\bibitem[Feng(2009)]{5319235}
L.~Feng.
\newblock Conplan: A context-aware ad-hoc meeting plan program.
\newblock In \emph{Symposia and Workshops on Ubiquitous, Autonomic and Trusted
  Computing}, pages 228--233, 2009.

\bibitem[Kleinberg and \'{E}va Tardos(2013)]{Kleinberg2013}
Jon Kleinberg and \'{E}va Tardos.
\newblock \emph{{Algorithms Design}}.
\newblock Pearson-Addison Wesley, 2013.
\newblock URL \url{http://www.cs.princeton.edu/~wayne/kleinberg-tardos}.

\bibitem[Xhafa et~al.(2016)Xhafa, Palou, Caballe, and Barolli]{7791872}
F.~Xhafa, D.~Palou, S.~Caballe, and L.~Barolli.
\newblock On sharing and synchronizing groupware calendars under android
  platform.
\newblock In \emph{International Conference on Complex, Intelligent, and
  Software Intensive Systems (CISIS)}, pages 122--129, 2016.

\bibitem[Siemoneit(2015)]{Siemoneit2015}
Andreas Siemoneit.
\newblock Navigation.
\newblock \emph{Nautisches Lexikon}, 2015.
\newblock URL \url{http://nautisches-lexikon.de}.

\bibitem[{Geo Midpoint}(2007)]{GeoMidpoint2007}
{Geo Midpoint}.
\newblock Geographic midpoint.
\newblock 2007.
\newblock URL \url{http://geomidpoint.com}.

\bibitem[Chaudhuri et~al.(1996)Chaudhuri, Garg, and Ravi]{Chaudhuri1996}
Shiva Chaudhuri, Naveen Garg, and R.~Ravi.
\newblock {Generalized k-Center Problems}.
\newblock \emph{Information Processing Letters}, 1996.

\bibitem[Francis et~al.(1991)Francis, McGinnis, and White]{Francis1991}
Richard~L. Francis, F.~McGinnis, and John~A. White.
\newblock \emph{{Facility Layout and Location: An Analytical Approach}},
  volume~2.
\newblock Pearson, 1991.

\end{thebibliography}

\end{document}